# CFLMEC: Cooperative Federated Learning for Mobile Edge Computing

Xinghan Wang+, Xiaoxiong Zhong+, ||, *, Yuanyuan Yang#, Tingting Yang||

+School of Cyber Science and Engineering, Southeast University, Nanjing, 211189, P. R. China

|| Peng Cheng Laboratory, Shenzhen, 518000, P. R. China

#Department of Electrical and Computer Engineering, Stony Brook University, Stony Brook, NY11794, USA

*Corresponding author: Xiaoxiong Zhong, email: xixzhong@gmail.com

*Abstract-* We investigate a cooperative federated learning framework among devices for mobile edge computing, named CFLMEC, where devices co-exist in a shared spectrum with interference. Keeping in view the time-average network throughput of cooperative federated learning framework and spectrum scarcity, we focus on maximize the admission data to the edge server or the near devices, which fills the gap of communication resource allocation for devices with federated learning. In CFLMEC, devices can transmit local models to the corresponding devices or the edge server in a relay race manner, and we use a decomposition approach to solve the resource optimization problem by considering maximum data rate on sub-channel, channel reuse and wireless resource allocation in which establishes a primal-dual learning framework and batch gradient decent to learn the dynamic network with outdated information and predict the sub-channel condition. With aim at maximizing throughput of devices, we propose communication resource allocation algorithms with and without sufficient sub-channels for strong reliance on edge servers (SRs) in cellular link, and interference aware communication resource allocation algorithm for less reliance on edge servers (LRs) in D2D link. Extensive simulation results demonstrate the CFLMEC can achieve the highest throughput of local devices comparing with existing works, meanwhile limiting the number of the sub-channels..

*Index Terms – federated learning*; mobile edge computing

## I. INTRODUCTION

With the great development of information and communications technology (ICT), more and more mobile devices are connected, which will need more bandwidth and bring a challenge for the capacity of computing and battery for mobile devices, if we exploit the cloud computing manner for them, it will have a high resource consumption and a high latency. Mobile Edge Computing (MEC) is a new promotion technology that extend the computing and storage at network edge, providing timely and reliable services and efficient bandwidth utilization [1]. On the other hand, mobile devices will generate huge amounts of data with privacy-sensitive in nature at the edge network. However, in this scenario, devices should share their own data to the connected server. Federated Learning (FL) [2] is a promising solution to solve such difficult problem. which can allow devices to build a consensus learning model with a collaborative and manner while preserving all training data on these devices. Each device can send the learning model to the server with its gradient and they are aggregated and feedback by the server. However, when mobile devices exploit an uncooperative training strategy, it is hard to improve the communication efficiency while updating model during aggregation. Hence, a challenging issue in FL is how devices cooperate to build a high-quality global model with considering communication resource allocation.

FL with resource allocation in MEC is a promising scheme for resource management in intelligent edge computing, improving resource utilization and preserving data privacy. Cooperative federated learning with resource optimization in an adaptive manner for MEC will brings some challenging issues. How could we design an efficient resource optimization framework for cooperative FL and how could we guarantee the optimal value to a resource management scheme and performance optimality given cooperative FL?

To answer these questions, we propose a cooperative federated learning framework for the MEC system, named CFLMEC, which mainly considers maximum data rate on sub-channel, channel reuse and wireless resource allocation. In CFLMEC, devices can transmit local models to the corresponding devices or the edge server in a relay race manner. The contributions of this article are as follows:

1) In order to make use of resource, we propose a cooperative federated learning for MEC, whose goal is maximizing the admission data to the edge server or the near devices. In CFLMEC, we use a decomposition approach to solve the problem by considering maximum data rate on sub-channel, channel reuse and wireless resource allocation in which establish a primal-dual learning framework and batch gradient decent to learn the dynamic network with outdated information and predict the sub-channel condition.
2) In CFLMEC, devices can transmit local models to the corresponding devices or the edge server in a relay race manner, which aims at maximizing throughput of

This work was supported by the National Natural Science Foundation of China (Grant Nos. 61802221, 61802220), and the Natural Science Foundation of Guangxi Province under grant 2017GXNSFAA198192, and the Key Research and Development Program for Guangdong Province 2019B010136001, the Peng Cheng Laboratory Project of Guangdong Province PCL2018KP005 and PCL2018KP004.

devices. To achieve this goal, we propose communication resource allocation algorithms with and without sufficient sub-channels for strong reliance on edge servers (SRs) in cellular link, and interference aware communication resource allocation algorithm less reliance on edge servers (LRs) in D2D link.
3) We present a new proactive scheduling policy, which allows an edge server to select the SRs and assigns sub-channels based on its sub-channel condition (we can see the details in Algorithm 4), outdated information from SRs (we can see the details in Algorithm 2), instantaneous information from SRs (we can see the details in Algorithm 1). For efficient sub-channels utilization, we assume sub channels reuse such that a sub-channel can be shared by at most two devices simultaneously. We need find a pair (LRS, SRs) for SRs with the same sub-channels and select a trans-mission power for LRs (we can see the details in Algorithm 3).
4) We conduct extensive experiments to evaluate the performance of the CFLMEC. With the numerical results, we show that the proposed method can achieve a higher throughput.

The remainder of this paper is organized as follows. Section II gives the related work. The detailed descriptions of CFLMEC will presented in Section III. We give the performance evaluation of the CFLMEC is in Section IV and make a conclusion for the paper in Section V.

## II. RELATED WORK

As a promising machine learning technique, federated learning based wireless network performance optimization has been attracted more attentions recently due to its good trade-off in data privacy risks and communication costs.

Most of existing works about FL in wireless networks mainly focus on resource allocation and scheduling. Dinh *et al*, [3] proposed the FEDL framework, which can handle heterogeneous mobile device data with only assumption of strongly convex and smooth loss functions. In FEDL, it exploits different models updating methods for local model and global model, which is based on corresponding computation rounds. And they implement FEDL for resource allocation optimization in wireless networks with heterogeneous computing and power resources. Ren *et al*. [4] mainly focused on federated edge learning with gradient averaging over selecting devices in each communication round, which exploits a novel scheduling policy with considering two types diversities about channels and learning updates. Yang *et al*., [5] studied three scheduling policies of federated learning (FL) in wireless networks: random scheduling, round robin, and proportional fair, and exploited a general model that accounts for scheduling schemes. Chen *et al.*, [6] studied the joint optimization problem that including device scheduling, learning, and resource allocation: which minimizes the FL loss function with transmission delay constrains. Ding *et al*. [7] presented a new server's optimal multi-dimensional contract-theoretic approach based incentive mechanism design with considering training cost and communication delay. In the meanwhile, they analyze the impact of information asymmetry levels on server's optimal strategy and minimum cost. Xia *et al*. [8] formulated a client scheduling problem as an optimization problem: minimizing the whole training time consumption, which includes transmission time and local computation time in both ideal and non-ideal scenarios. And then they used a multi-armed bandit based scheme to learn to scheduling clients online in FL training without knowing wireless channel state information and dynamics of computing resource usage of clients. Aiming at accelerating the training process in FL, Ren *et al*. [9] formulate a training acceleration optimization problem as a maximizing the system learning efficiency problem, in the CPU scenario or GPU scenario, which jointly considers batch size selection and communication resource allocation. Pandey *et al*. [10] proposed a novel incentive based crowd-sourcing framework to enable FL, in which exploited a two-stage Stackelberg game model to maximize the utility of the participating clients and MEC server interacting. Considering probabilistic queuing delays, Samarakoon *et al*. [11] studied the problem of joint federated learning based power and resource allocation in vehicular networks, minimized power consumption of vehicular users and estimated queue lengths distribution using by Lyapunov optimization in wireless links communication delays. Shi *et al.* [12] formulated the problem of joint bandwidth allocation and devices scheduling as maximize the convergence rate problem, which is to capture the long-term convergence performance of FL.

For optimizing FL mechanism in wireless networks, some proposals have been presented. In order to optimize the expected convergence speed, Nguyen *et al*. [13] proposed a fast convergent federated learning algorithm, which can deal with the heterogeneity of computation and communication of devices by adapting the aggregations based to the device's contributions for updating. Mills *et al*. [14] presented an adapting FedAvg to exploit a distributed manner of Adam optimization and the novel compression techniques, which can greatly reduce the number of rounds to convergence. Guo *et al*. [15] proposed a novel analog gradient aggregation in wireless networks, which can improve gradient aggregation quality and accelerate convergence speed. Wang *et al*. [16] studied the problem of learning model parameters in the FL framework analyzed the convergence bound of distributed gradient descent from a theoretical perspective, which is based on the proposed control algorithm for minimizing the loss function with a resource budget constrain.

To the best of our knowledge, there are few works about decentralized FL in wireless networks. Luo *et al*. [17] presented a novel hierarchical federated edge learning (HFEL) framework in which model aggregation is partially migrated to edge servers from the cloud. In HFEL, they studied the resource optimization problem formulated as a global cost minimization, and decomposed it into two sub problems: resource allocation and edge association. Savazzi *et al*. [18] proposed a novel device cooperation FL framework based on the iterative exchange of both model updates and gradients, which can improve convergence and minimize the number of communication rounds in the D2D network. However, they did not completely transmit local models in a cooperative manner, e.g., they can only transmit local models to an edge server, or only transmit local models to a device without considering channel allocation.

All of the above-mentioned existing works of federated learning focused on designing learning algorithm to improve training performance or maximizing network performance, the cooperative federated learning issue among devices is still under-explored, which will cause a poor system performance for the FL based MEC system. Hence, how to design an efficient cooperative federated learning framework that device not only transmit local model to an edge server but also transmit local models to its near devices in a relay race manner, with considering resource allocation for MEC is a challenging issue. This paper aims to propose a solution to address this problem.

## III. MODEL FOR CFLMEC

In this section, we will describe the architecture model, mathematical model and communication model for the proposed cooperative federated learning, CFLMEC.

### 1. Cooperative federated learning architecture model

In this paper, we consider a cooperative federated leaning system with an edge server and multiple local devices, The set of local devices denoted as $M = \{1, 2, 3..., M\}$.

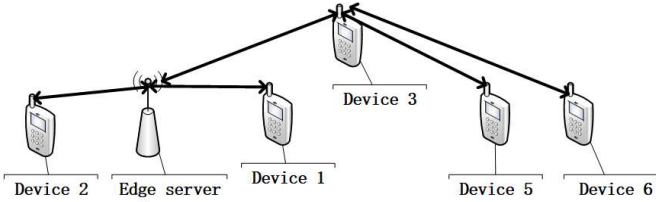

Fig. 1. Cooperative federated learning architecture.

In the proposed architecture, local devices are divided into two types: local devices with less reliance on edge server (LRs) and local devices with strong reliance on edge server (SRs). The set of LRs $K \subseteq M = \{1, 2, 3..., K\}$ consists of all such local devices which can not be directly connect to edge server due to harvested energy limitations and a high transmission delay. The set $H \subseteq M = \{1, 2, 3..., H\}$ of SRs consists of all such local devices which can be connected to edge server. Thus, cooperative federated learning requires LRs to send their local models to the near SRs, then the SRs must both aggregates the local models received from LRs and train its local model. Finally, the BS (edge server) aggregates models received from SRs and transmits it to the associated devices. For example, as shown in Fig.1, the device 5 and device 6 send local model to the device 3, the device 3 can be consider as SRs, the device 5 and device 6 can be considered as LRs, then device 3 trains its local model using gradient decent and aggregates local model from device 5 and device 6 while the edge server aggregates the models from device 3.

Due to limited harvested energy and high transmission delay, a LRs can transmit local model to one of the SRs. To represent the local devices association, we introduce a binary indicator variable $x_{kh} \in \{0,1\}$, where $k \in K$ and $h \in H$, and define the device profile as $x \in \{x_{kh} \mid k \in K, h \in H\}$.

### 2. Mathematical demonstration

In this subsection, we introduce the leaning process. As shown in Fig. 2, the LRs are allocated to an SRs, and the edge server collectively learns the global model with the help of the SRs.

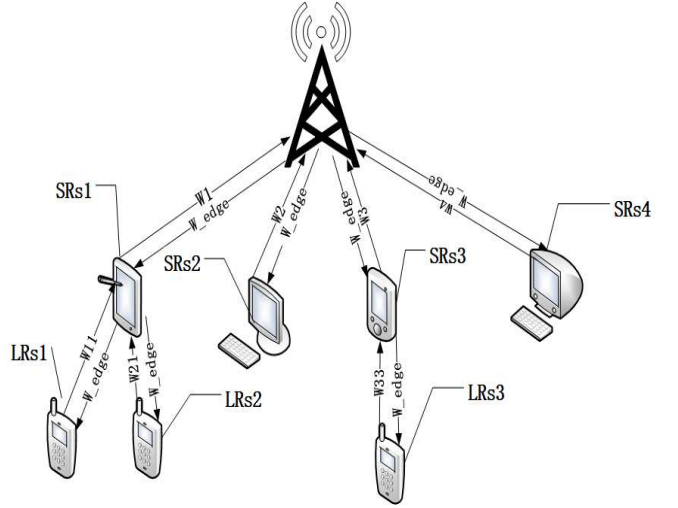

Fig. 2. The cooperative federated learning weight update.

Each local device $m$ collects a matrix $X_m = \{x_{m1}, x_{m2}, ..., x_{mL_m}\}$, where $L_m$ is the number of the samples collected by device $m$. The output data vector for training cooperative federated learning of local device $m$ is $Y_m = \{y_{m1}, y_{m2}, ..., y_{mL_m}\}$. Let $W_m$ denote the parameters related to model that is trained by $X_m$ and $Y_m$. We refer to the dataset of each device by $D_m$. Upon a specific assignment, each SRs can collect models from the near LRs and the edge server can only receive the models from the SRs. The aggregated dataset of each SRs is $D_h^{aggregate} = \left| D_h + \bigcup x_{kh} D_k \right|$, and the aggregated dataset of the edge server $D_{aggregate}^{edge\ server} = \left| \bigcup D_h^{aggregate} \right|$. The objective of the overall cooperative learning process is to converge to $w^*$ which solves the following problem:

$$\arg\min_w F(w) = \frac{1}{N} \sum_{m=1}^{M} \sum_{n=1}^{L_m} f(w, x_{mn}, y_{mn}) \quad (1)$$

where $N = \sum_{m=1}^{M} L_m$ is the total number of the dataset belonged to the devices. For the $k$-th local devices, the local parameters at time slot $t$ are optimized as follow:

$$w_k(t) = w_k(t-1) - \delta \nabla_k F_k(w_k) \quad (2)$$

Specially, in cooperative federated learning, the weights are synchronized across LRs belonging to the SRs. Hence, at time

slot $t$, the parameters of an SRs aggregation are:

$$w_h^{aggregate}(t) = \sum_{k=1}^{K} \frac{x_{kh} D_k}{D_h^{aggregate}} w_k(t) + \frac{D_h}{D_h^{aggregate}} w_h(t) \quad (3)$$

Similarly, at edge server, at time slot $t$, the weights are averaged across all SRs in edge:

$$w_{edge}(t) = \sum_{h=1}^{H} \frac{D_h^{aggregate}}{D_{aggregate}^{edge\ server}} w_h^{aggregate}(t) \quad (4)$$

## 3. Communication model

As shown in Fig.3, we introduce the links and local devices association in our network. Each device transmits its trained local model to its connected devices or the edge server via a shared wireless interface with $N$ sub-channels. We introduce links in the network as follows:

**Cellular link**: an SRs can transmit its aggregate models to the edge server or directly transmit its local model to the edge server (there is no LRs belonging to this SRs).

**D2D link**: a LRs can establish a direct D2D link with the nearest SRs within the maximum distance.

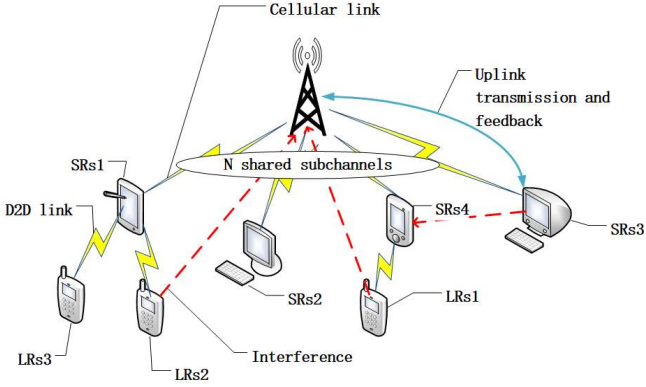

Fig. 3. An illustration of the cooperative federated learning with resource allocation.

For efficiently using sub-channels, we assume channels reuse such that a sub-channel can be shared by at most two devices simultaneously. Therefore, the reuse of sub-channel is allowed only to a cellular and a D2D link, and are not among the D2D links. Let $N = \{1,2,...,N\}$ denote $N$ channels, and thus the available bandwidth $B$ is divided into $N$ orthogonal sub-channels. In our case, a D2D link reuses the sub-channel of a cellular link, so we must consider the interference. And the SINR of cellular link can be expressed as:

$$\gamma_h^n = \frac{p_{h\_edge}^n h_{h\_edge}^n}{N_0 + \sum_{\substack{h'=1 \\ h' \neq h}}^{H} \sum_{k=1}^{K} x_{kh'} p_{kh'}^n h_{k\_edge}^n} \quad (5)$$

where $p_{h\_edge}^n$ is the transmission power from SRs to edge and $h_{h\_edge}^n$ denote the channel gain between SRs and edge server on sub-channel $n$. Let $N_0$ denote the noise power and $h_{k\_edge}^n$ is the channel gain of interference link between LRs and the edge server.

Therefore, the SINR of D2D link when it reuses sub-channel $n$ can be expressed as:

$$\gamma_k^n = \frac{x_{kh}^n p_{kh}^n h_{kh}^n}{N_0 + \sum_{\substack{h'=1 \\ h' \neq h}}^{H} p_{h'\_server}^n h_{h'\_h}^n} \quad (6)$$

Then the data rate of each device $m$ on sub-channel $n$ can be expressed as:

$$r_m^n = \bar{B} \log_2\left(1 + \gamma_m^n\right) \quad (7)$$

where $\bar{B}$ is the bandwidth per sub-channels. The total data rate of each device $m$ can be defined as:

$$R_m = \sum_{n=1}^{N} r_m^n \quad (8)$$

We further discuss the state of local devices. Let $S(t) = \{S_m(t)\}$ collect the state of all local devices at time slot $t$, where $S_m(t) = \{A_m(t), R_m(t)\}$ is the state of the device $m$. $A_m(t)$ is the data size collected by device $m$ at time slot $t$ corresponding to the data size of parameter $w_m(t)$. $R_m(t)$ is the instantaneous capacity of sub-channels at device $m$.

Having all the local devices to report their state to the BS at each time slot may be not easy, so we divide the situation into two parts: one part is that there are enough sub-channels for the local devices, hence the local device can select best quality sub-channels. We can assign one sub-channel to a local device based on maximum data rate. As there is only one sub-channel initially, maximum power is allocated to local device. We then can assign the remaining sub-channel to local device, so the local device can have more than one sub-channels. Another part is that we consider that the local devices can be in large scale network. So we meet the challenge where there are not enough sub-channels, which means that we cannot assign the sub-channels to the local devices immediately. For example, the local device generates the data in time slot $T_m(t_0)$ and local device cannot send the data to the edge server or its near devices immediately, since the number of devices can be dramatically smaller that the number of sub channels. In this scenario, the edge server can only schedule the devices based on outdated state of local devices. Let $T_m(t)$ be the time that we assign the sub-channel to device $m$. So each device maintains data from time $T_m(t_0)$ to $T_m(t)$ in data queue.

At time slot $t$, device $m$ can be admitted to transmit the queue data to edge server or near SRs. The admission data $a_m(t)$ meets constrains s follow:

$$\overline{a_m(t) - R_m(t)} \leq 0 \\ 0 \leq a_m(t) \leq A_m(t) \quad (9)$$

where $\overline{x(t)} = \lim_{T \to \infty} \frac{1}{T} \sum_{t=0}^{T-1} E[x(t)]$ indices that average of a process. The first constrain denotes admission data at device is more than the data rate of device. The second constrain denotes the admission data at device, in which there is no more than the arrived data at device during time slot $t$.

Let $\rho(t) = \{\rho_m^n(t)\}$ denote the device schedule decision at time slot $t$. $\rho_m^n(t)=1$ denotes the device $m$ selects channel $n$ at time slot $t$ Otherwise $\rho_m^n(t)=0$

## IV. PROBLEM FORMULATION

In this work, we aim to maximize the admission data of devices which transmit their models to the edge server or the SRs. The purpose is to maximize the time-average network throughput of cooperative federated learning framework, which is based on QoS aware communication resource allocation with sufficient sub-channels, batch gradient descent and primal-dual predict learning without sufficient channels, and the optimal schedule with a learned online method.

Based on the system model in Section III, we formulate communication resource allocation scheme as follows:

$$P1: \max_{Q(t)} \overline{\sum_{m=1}^{M} a_m(t)}$$

$$s.t. \quad \overline{a_m(t) - R_m(t)} \le 0 \quad (10\text{-}1)$$
$$0 \le a_m(t) \le A_m(t) \quad (10\text{-}2)$$
$$\sum_{k=1}^{K} \sum_{h=1}^{H} \rho_k^n(t) \rho_h^n(t) \le 2 \quad (10\text{-}3)$$
$$\sum_{h=1}^{H} x_{kh} = 1 \quad (10\text{-}4)$$
$$\sum_{h=1}^{H} \rho_h^n(t) \le 1 \quad (10\text{-}5) \quad (10)$$
$$\sum_{k=1}^{K} \rho_k^n(t) \le 1 \quad (10\text{-}6)$$
$$\sum_{n=1}^{N} \rho_k^n(t) \ge 1 \quad (10\text{-}7)$$
$$\sum_{m=1}^{M} \sum_{n=1}^{N} P_m^n \le P_m^{\max} \quad (10\text{-}8)$$

where $Q(t) = \{a_m(t), \rho_m(t)\}$ denotes the data admission and schedule from all device across all sub-channels at time slot $t$.

Here, the objective function aims to maximize the data admission of network. The constrain in (10-1) implies that the admission data must not exceed the maximum data rate on sub-channel. Whereas (10-2) shows admission data is between zero and collected data at device at time slot. In (10-3), the sub-channel can be shared by at most two links and only a cellular and a D2D link are allowed to reuse the sub-channel $n$. The constrain in (10-4) shows that a LRs can only connect to a SRs. The constraints in (10-5) and (10-6) present the sub-channel condition, where (10-5) implies that a sub-channel can shared by one cellular link and no more than one cellular link, such as two cellular links is not allowed in a sub-channel, and (10-6) implies that a sub-channel can shared by one D2D link and no more than one D2D link, such as two D2D links are not allowed in a sub-channel. The (10-7) implies that a SRs can occupy more than one sub-channels. The (10-8) implies that a local device the transmission power cannot exceed the maximum transmission power of local device.

We discuss the situation with two parts. One part is that communication resource allocation with sufficient sub-channels. Another part is the communication resource allocation without sufficient sub-channels in large scale federated learning.

**Theorem 1**: The objection function in **P1** is strong convex.

**Proof.** Let $P1: \max_Q \overline{\sum_{m=1}^{M} a_m(t)} = \lim_{T \to \infty} \frac{1}{T} \sum_{t=0}^{T-1} \sum_{m=1}^{M} a_m(t)$ denote the time average objective. The Hessian Matrix of **P1** is positive and $\frac{\partial \overline{\sum_{m=1}^{M} a_m(t)}}{\partial \tau_{m_1}(t_1) \tau_{m_2}(t_2)} = 0$, if $m_1 \ne m_2$ or $t_1 \ne t_2$. Therefore, the convexity of the objective function is confirmed.

This completes the proof.

**Theorem 2**: Given an edge server and a set of SRs and LRs, we can have the divergence, $w_h^{aggregate}(t)$, which is equal to the weights reached by using centralized gradient decent on the $h$-th aggregated at time slot $t$:

$$w_h^{aggregate}(t) = w_h^{aggregate}(t-1) - \delta \nabla F(w_h^{aggregate}(t-1))$$

**Proof.** From equitation (3), we can have

$$w_h^{aggregate}(t) = \sum_{k=1}^{K} \frac{x_{kh} D_k}{D_h^{aggregate}} w_k(t) + \frac{D_h}{D_h^{aggregate}} w_h(t)$$

$$= \sum_{k=1}^{K} \frac{x_{kh} D_k}{D_h^{aggregate}} (w_k(t-1) - \delta \nabla F(w_k(t-1))$$

$$+ \frac{D_h}{D_h^{aggregate}} (w_h(t-1) - \delta \nabla F(w_h(t-1))$$

$$= \sum_{k=1}^{K} \frac{x_{kh} D_k}{D_h^{aggregate}} w_k(t-1) + \frac{D_h}{D_h^{aggregate}} w_h(t-1)$$

$$- (\sum_{k=1}^{K} \frac{x_{kh} D_k}{D_h^{aggregate}} \delta \nabla F(w_k(t-1)) + \frac{D_h}{D_h^{aggregate}} \delta \nabla F(w_h(t-1))$$

(11)

Since

$$\sum_{k=1}^{K} \frac{x_{kh} D_k}{D_h^{aggregate}} w_k(t-1) + \frac{D_h}{D_h^{aggregate}} w_h(t-1) = w_h^{aggregate}(t-1) \quad (12)$$

and

$$\nabla F(w_h^{aggregate}(t-1)) = \sum_{k=1}^{K} \frac{x_{kh} D_k}{D_h^{aggregate}} \nabla F(w_k(t-1)) + \frac{D_h}{D_h^{aggregate}} \delta \nabla F(w_h(t-1))$$

(13)

Submitting (13) and (12) into (11), we can have:

$$w_h^{aggregate}(t) = w_h^{aggregate}(t-1) - \delta \nabla F(w_h^{aggregate}(t-1))$$

This completes the proof.

**1. Communication resource allocation with sufficient sub-channels for SRs in cellular link.**

In the considered D2D assisted cooperative federated learning in cellular network. After assigning the sub-channels to local devices based on the maximum data rate, we can assign the remaining sub-channel to device whose QoS is met the smallest value, we can select best sub-channel for the weakest cellular link. We perform communication resource allocation in **Algorithm 1**, which can improve the data admission in cellular link for SRs.

In **Algorithm 1**, we first consider that a sub-channel should be assigned to an SRs based on the maximum data rate such that $(h^*, n^*) = \arg\max_{h \in H_c, n \in N}(\frac{r_h^n(t)}{\sum_{h'=1}^{H} r_{h'}^n(t)/|H_c|})$. Due to the fact that we have the sufficient sub-channels, we can make some supplement to some SRs that are allowed slow admission data until that sub-channels are fully utilized. For that reason, that we assign one sub-channels to a SRs at first, maximum power is allowed to the SRs. We then assign the remaining the sub-channels to the SRs according to the $h' = \arg\min_{h \in H_c}(\frac{a_h(t)}{\sum_{h'=1}^{H} a_{h'}(t)/|H_c|})$. We assign more than one channel to the weakest cellular link and improve the size of admission data.

---

**Algorithm 1**: QoS aware communication resource allocation for SRs in cellular link with sufficient sub-channels

**Input:** $N'$, $Q$, $K'$, $H'$, $M' = K' \cup H'$, $h_{h\_edge}^n$

**Output:** $\rho, a, P$

1: Initialize $H = H'$, $num_h^{sub\_channel} = 0$, $R_h = 0$, $N = N'$, $H_c = H'$
2: **while** $H \neq \emptyset$ **do**
3: Find $(h^*, n^*) = \arg\max_{h \in H_c, n \in N}(\frac{r_h^n(t)}{\sum_{h'=1}^{H} r_{h'}^n(t)/|H_c|})$
4: Set $\rho_h^n(t)=1$ and update $num_h^{sub\_channel} = num_h^{sub\_channel} + 1$
5: Set $p_{h\_edge}^n(t) = p_{h\_edge}^{max}$ and $a_h(t) = r_h^n(t)$
6: Update $H = H - \{h^*\}$
7: Update $N = N - \{n^*\}$
8: **end while**
9: **while** $N \neq \emptyset$
10: Find $h' = \arg\min_{h \in H_c}(\frac{a_h(t)}{\sum_{h'=1}^{H} a_{h'}(t)/|H_c|})$
11: Find $n^* = \arg\max_{n \in N}(r_{h'}^n(t))$
12: Set $\rho_{h'}^n(t)=1$ and update $num_{h'}^{sub\_channel} = num_{h'}^{sub\_channel} + 1$
13: Update power allocation $p_{h'\_edge}^n(t) = p_{h'\_edge}^{max} / num_{h'}^{sub\_channel}$
14: Update $a_{h'}(t) = a_{h'}(t) + r_{h'}^n(t)$
15: Update $N = N - \{n^*\}$
16: **end while**

---

Then, we discuss the complexity of the **Algorithm 1**. In the first, there are $H$ iterations for initial sub-channel to the SRs. The search for an optimal pair is $O(HN)$, thus the complexity of initial sub-channel assignment is $O(H^2N)$. For the remaining sub-channels allocation, the complexity is $O(H(N-H)^2)$. So the whole complexity of **Algorithm 1** is $O(H^2N)+O(H(N-H)^2)$.

**2. Communication resource allocation without sufficient sub-channels for SRs in cellular link.**

In this case, we consider that there are not enough sub-channels for SRs, so the problem is challenging with the following features:

The edge server can collect information of SRs delay by $T_h(t)-T_h(t_0)+1$ time slots where $T_h(t_0)$ is the time that device will begin to generate data and $T_h(t)$ be the time that we assign the sub-channel to device. The edge server makes the scheduling decision in the absence of the devices state. To address the challenge, we develop batch gradient descent and primal-dual predict learning without sufficient channels. The objective can be transformed as:

$$P2 \quad \min_{Q(t)} -\overline{\sum_{h=1}^{H} a_h(t)}$$

So the Lagrangian of P2 can be reformulated by:

$$L(Q(t), S(t), \lambda(t)) = -\sum_{h=1}^{H} a_h(t) + \sum_{h=1}^{H} \lambda_h[a_h(t) - \sum_{n=1}^{N} \rho_h^n(t)r_h^n(t)] \quad (14)$$

Given the convex object and constraints, P2 can be reformulated by primal-dual predict learning to its dual problem:

$$\min L(Q(t), S(t), \lambda(t)) = \max G(\lambda(t)) \quad (15)$$

The duality gap can be diminished by finding the optimal multipliers to maximize the dual lagrangian. Next, we need to find the optimal primal variables $\min_{Q(t)} L(Q(t), S(t), \lambda(t))$ and dual multipliers $\max_{\lambda(t)}(\min_{Q(t)} L(Q(t), S(t), \lambda(t)))$

Batch gradient decent: The edge server can update multipliers for the scheduled device according to:

$$\lambda_h(t+1) = [\lambda_h(t) + \varepsilon(\sum_{t=T_h(t_0)}^{T_h(t)} a_h(t) - \sum_{n=1}^{N} \rho_h^n(T_h(t))r_h^n(T_h(t)))]^+ \quad (16)$$

Here, the above equation can be regarded as batch gradient decent with outdated information of SRs. By using the multipliers, we can get the optimal primal variables. The update can be given by:

$$Q^*(t) = \arg\min_{Q(t)} L(Q(t), S(t), \lambda(t)) \quad (17)$$

**Algorithm 2**: QoS aware communication resource allocation for SRs in cellular link without sufficient sub-channels
---
**Input:** $N', Q, K', H', M' = K' \cup H', h_{h\_edge}^n$
**Output:** $\rho, a, P$
1: **Initialize** $H = H', \lambda(t) = 0, N = N'$
2: At $T_h(t_0)$, calculate t $a_h(t)$ using by primal-dual predict learning.
3: At $T_h(t)$, edge server receives the state from device
4: Update multipliers $\lambda(t+1)$ according to (16).
5: If $\rho_h^n(t)=1$
6: Calculate $\rho_h^n(t)$ at edge server.
7  Set $p_{h\_edge}^n(t) = p_{h\_edge}^{\max}$
8: Update devices state $S(t)$
---

Then, we discuss the complexity of the **Algorithm 2**. Each SRs hold its multiplier $\lambda_h(t)$, and optimize its admission data $a_h(t)$, the complexity of SRs is $O(1)$. The edge server calculate the $\lambda_h(t+1)$, and update $\lambda_h(t+1)$ according to the batch gradient decent. The edge server assign the sub-channels to SRs and calculate $\rho_h^n(t)$. Hence, the complexity of edge server is $O(N^2 H)$.

**Theorem 3**: Given the multipliers for the scheduled device $\lambda_h(t)$, the $Q(t)$ can be given by:

$$a_h(t) = \begin{cases} 0, & \text{if } \alpha - \alpha\lambda_h(t) + a_h(t-1) \leq 0 \\ A_h(t), & \text{if } \alpha - \alpha\lambda_h(t) + a_h(t-1) \geq A_h(t) \\ \alpha - \alpha\lambda_h(t) + a_h(t-1), & \text{otherwise} \end{cases}$$

$$\rho_h^n(t) = \begin{cases} 1 & h = h' \\ 0 & h \neq h' \end{cases} \text{ and}$$

$$\underset{h'}{\arg\min} \sum_{h'=1}^{H} \frac{(\rho_{h'}^n)^2}{2\alpha} + (-\lambda_{h'}(t)r_{h'}^n(t-1) - \frac{\rho_{h'}^n(t-1)}{\alpha})\rho_{h'}^n \quad \forall n \in N$$

**Proof**: To get the optimal primal variables, we need to find the point which has the minimum distance with the $Q(t-1) - \alpha\nabla L_Q(Q(t-1), \lambda(t), S(t-1))$, thus the distance problem can be reformulated as given by:

$$Q(t) = \arg\min \nabla L_Q(Q(t-1), \lambda(t), S(t-1))(Q - Q(t-1)) + \frac{\|Q - Q(t-1)\|^2}{2\alpha} \quad (18)$$

The above equation can be transformed as:

$$\begin{cases} \rho(t) = \arg\min \nabla L_\rho(Q(t-1), \lambda(t), S(t-1))(Q - Q(t-1)) \\ \qquad + \frac{\|\rho - \rho(t-1)\|^2}{2\alpha} \\ a(t) = \arg\min \nabla L_a(Q(t-1), \lambda(t), S(t-1))(Q - Q(t-1)) \\ \qquad + \frac{\|a - a(t-1)\|^2}{2\alpha} \end{cases} \quad (19)$$

Optimal admission data: The SRs can solve the following problem for the admission data:

$$\begin{cases} \nabla L_a(Q(t-1), \lambda(t), S(t-1))(Q - Q(t-1)) + \frac{\|a - a(t-1)\|^2}{2\alpha} \end{cases}$$

$$= \sum_{h=1}^{H} [\frac{a_h^2}{2\alpha} + (-1 + \lambda_h(t) - \frac{a_h(t-1)}{\alpha})a_h]$$

$$+ \sum_{h=1}^{H} [\frac{a_h^2(t-1)}{2\alpha} + (1 - \lambda_h(t))a_h(t-1)] \quad (20)$$

$$\begin{cases} \nabla L_\rho(Q(t-1), \lambda(t), S(t-1))(Q - Q(t-1)) \\ + \frac{\|\rho - \rho(t-1)\|^2}{2\alpha} \end{cases}$$

$$= \sum_{h=1}^{H}\sum_{n=1}^{N} [\frac{(\rho_h^n)^2}{2\alpha} + (-\lambda_h(t)r_h^n(t-1) - \frac{\rho_h^n(t-1)}{\alpha})\rho_h^n] \quad (21)$$

$$+ \sum_{h=1}^{H}\sum_{n=1}^{N} [(\lambda_h(t)r_h^n(t-1))\rho_h^n(t-1)] + \sum_{h=1}^{H}\sum_{n=1}^{N} [\frac{(\rho_h^n(t-1))^2}{2\alpha}]$$

The above equations refer to the variables $a_h(t)$ and $\rho_h^n(t)$. The above equations can be suppressed as:

$$\begin{cases} \varpi(a) = \sum_{h=1}^{H} [\frac{a_h^2}{2\alpha} + (-1 + \lambda_h(t) - \frac{a_h(t-1)}{\alpha})a_h] \\ \eta(\rho) = \sum_{h=1}^{H}\sum_{n=1}^{N} [\frac{(\rho_h^n)^2}{2\alpha} + (-\lambda_h(t)r_h^n(t-1) - \frac{\rho_h^n(t-1)}{\alpha})\rho_h^n] \end{cases} \quad (22)$$

The objective $\varpi(a)$ can get the optimal admission data:

$$\min \varpi(a) \quad \text{s.t } 0 \leq a(t) \leq A(t) \quad (23)$$

Then, we can get the optimal admission data:

$$a_h(t) = \begin{cases} 0, & \text{if } \alpha - \alpha\lambda_h(t) + a_h(t-1) \leq 0 \\ A_h(t), & \text{if } \alpha - \alpha\lambda_h(t) + a_h(t-1) \geq A_h(t) \\ \alpha - \alpha\lambda_h(t) + a_h(t-1), & \text{otherwise} \end{cases} \quad (24)$$

The objective $\eta(\rho)$ can get the sub-channel selection :

$$\min \eta(\rho) \quad \text{s.t } \rho \in \{0,1\} \quad (25)$$

The objective $\eta(\rho)$ can be decoupled between different sub-channels.

$$\underset{h'}{\arg\min} \sum_{h'=1}^{H} \frac{(\rho_{h'}^n)^2}{2\alpha} + (-\lambda_{h'}(t)r_{h'}^n(t-1) - \frac{\rho_{h'}^n(t-1)}{\alpha})\rho_{h'}^n \quad \forall n \in N$$

$$\rho_h^n(t) = \begin{cases} 1 & h = h' \\ 0 & h \neq h' \end{cases}$$

This completes the proof.

**Theorem 4**: We can conclude that $\lambda_h(t) \leq \lambda_h^{\max}$, where $\lambda_h^{\max} = (\varepsilon + \frac{1}{\alpha})A_h^{\max} + 1$.

**Proof**: if $-1+\lambda_h(t)-\frac{a_h(t-1)}{\alpha} \geq 0$, we can conclude that $\alpha-\alpha\lambda_h(t)+a_h(t-1) \leq 0$, hence, we have $a_h(t)=0$ according to (24), and $\lambda_h(t+1) \leq \lambda_h(t)$ according to (16). Hence, $\lambda_h(t) \leq \lambda_h^{max}$. If $-1+\lambda_h(t)-\frac{A_h^{max}}{\alpha} \leq 0$ and $a_h(t) \leq A_h^{max}$, we can obtain

$$\lambda_h(t+1) \leq \lambda_h(t)+\varepsilon A_h^{max} \leq 1+\frac{A_h^{max}}{\alpha}+\varepsilon A_h^{max}$$
$$=1+(\frac{1}{\alpha}+\varepsilon)A_h^{max}$$
$$=\lambda_h^{max}$$

This completes the proof.

### 3. Communication resource allocation for LRs in D2D link.

In initially, some SRs do not need to aggregate the model from LRs, because there are no LRs that belongs to SRs. These SRs can directly transmit its local model to edge server. At the same time, LRs need to share local model to the near SRs (these SRs cannot directly transmit its local model to edge server) in order to aggregate data.

To avoid degradation of weak cellular links of SRs, we should guarantee the admission data of SRs. We first range the SRs based on admission data. We then find a pair (LRS, SRs) for SRs with the same sub-channels and select a transmission power for LRs. In this paper we focus on the maximum transmission power for SRs, as shown in **Algorithm 3**, which is obviously to obtain its complexity with $O(K(H-1))$.

---

**Algorithm 3**: Interference aware communication resource allocation for LRs in D2D link

**Input**: $N', Q, K', H', M' = K' \cup H', h_{k\_edge}^n$

**Output**: $\rho, a, P$

1: **Initialize** $H = H', N = N', K = K'$
2: Range SRs that are in $H$ with a descending order of admission data
3: **For** $k = 1$ to $K$
4:   Find a pair $(h', h)$ with minimum $h_{h'\_h}^n$ where $x_{kh}=1$
5:   Assign $\rho_k^n(t)=1$ and $\rho_h^n(t)=1$
6:   Set $(P_{kh}^n)_1 = \dfrac{\gamma_k^n(N_0+\sum_{\substack{h'=1\\h'\neq h}}^H p_{h'\_server}^n h_{h'\_h}^n)}{x_{kh}^n h_{kh}^n}$
7:   Set $(P_{kh}^n)_2 = \dfrac{p_{h\_edge}^n h_{h\_edge}^n - \gamma_h^n N_0}{\gamma_h^n h_{h\_edge}^n}$
8:   Update $P_{kh}^n(t) = \min\{(P_{kh}^n)_1, (P_{kh}^n)_2\}$
9:   **if** $P_{kh}^n(t) > P_{h\_edge}^n$, **return.**
10:  **else** $\gamma_k^n = \dfrac{x_{kh}^n p_{kh}^n h_{kh}^n}{N_0+\sum_{\substack{h'=1\\h'\neq h}}^H p_{h'\_server}^n h_{h'\_h}^n}$
11:  Set $a_k(t) = r_k^n(t)$
12:  $K = K - \{k\}$
13:  $H = H - \{h\}$
14:  $N = N - \{n\}$
15: **end for**

---

Hence, the communication resource allocation for local devices is describe as following **Algorithm 4**.

---

**Algorithm 4**: Communication resource allocation for local devices

**Input**: $N', Q, K', H', M' = K' \cup H', h_{k\_edge}^n, h_{h\_edge}^n$

**Output**: $\rho, a, P$

1: **Initialize** $H = H', N = N', K = K'$, $num_h^{sub\_channel}=0, R_h=0, D_h$
2: Determine local devices association $X$
3: **for** $k=1$ to $K$
4:   **for** $h=1$ to $H$
5:     **if** $k$ is inside range $D_h$ **then**
6:       $x_{kh}=1$
7:     **end if**
8:   **end for**
9: **end for**
10: Based on number of sub-channels, we determine the communication resources for SRs in cellular link.
11: **if** $N > H$ the number of sub-channels is larger than the number of SRs **then**
12:   Executing **Algorithm 1**
13: **else**
14:   Executing **Algorithm 2**
15: **end if**
16: Allocate communication resources to LRs using **Algorithm 3**

---

## IV. PERFORMANCE EVALUATION

In this part, we evaluate the performance of the proposed CFLMEC framework. We establish the following parameters: we consider a network topology of 300 m × 300 m, which consists of one edge server, multiple are local device are randomly distributed. The maximum transmission power of mobile user set to 100 mW. The Rayleigh fading model is adopted for small scale fading. The bandwidth of edge server is 10 MHz. We set the network coverage radius of SRs as 50m. The channel gain is modeled as independent Rayleigh fading channel which incorporates the path loss and shadowing effects. The average channel capacity of the devices follows a uniform distribution within [0, 125] Kbps.

The number of sub-channels is 10, the data arrivals at device within a time slot is [0,40] Kbits. The numerical of baseline is offline optimum. Fig. 4 clearly reveals the change of network throughput with the different parameters $\varepsilon=0.001$, $\varepsilon=0.005$, and $\varepsilon=0.00025$ respectively. From this figure, $\varepsilon=0.00025$ is ranked in the first, $\varepsilon=0.005$ was far behind $\varepsilon=0.00025$, while the figure for 0.001 was the smallest compared with other parameters. We can see that the network throughput of the proposed approach increases with the growing number of devices.

Fig. 5 shows that the Lagrange multipliers of the Algorithm 2. It first increases under all different parameters ε, and then stabilizes at the same value over time. As the step size learning rate decreases from 0.002 to 0.0005, Algorithm 2 requires increasingly long convergence times to stabilize the system.

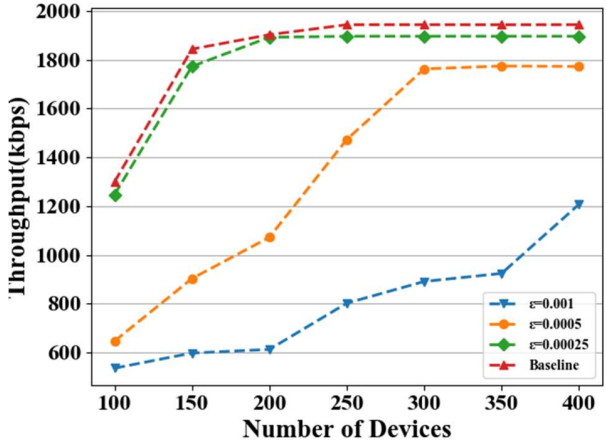

Fig. 4. Number of devices vs. throughput.

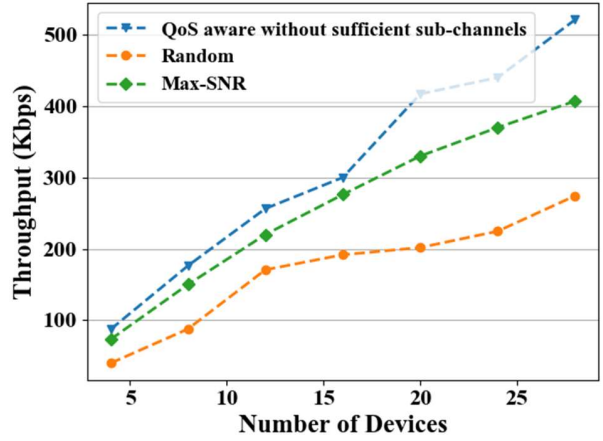

Fig. 7. Number of devices vs. throughput.

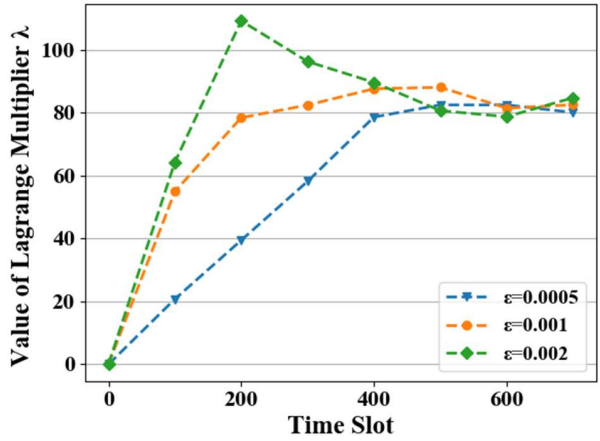

Fig. 5. Time slot vs. value of Lagrange multipliers.

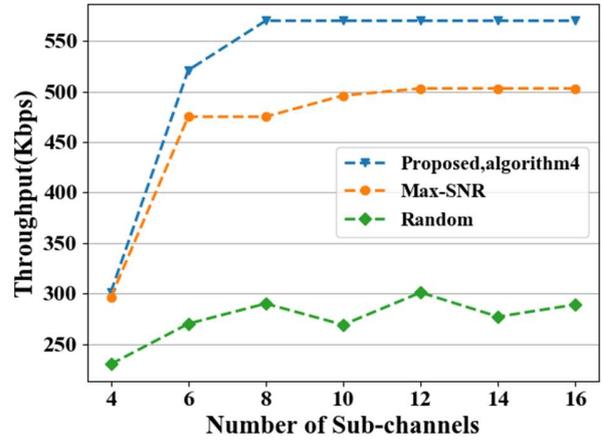

Fig.8. number of sub-channel vs. throughput.

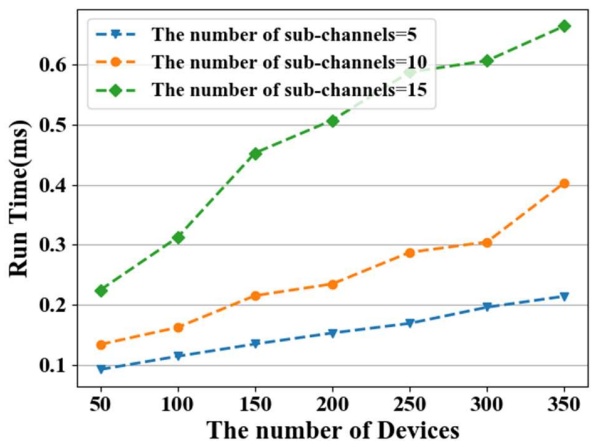

Fig. 6. Number of devices vs. run time.

Fig. 6. shows the runtime for different numbers of devices among different numbers of sub-channels. The learning rate is ε=0.00025. We can see that the runtime of the Algorithm 2 increases proportionally with the number of devices.

From the Fig.7, we can see that when the number of the devices increases from 4 to 28, the network throughput of three approaches increases, the values are from 87 Kbps to 512.322 Kbps for Algorithm 1+Algorithm 3, from 39.31 to 274.3 for Random, and from 73 Kbps to 407.382 Kbps for Max-SNR. Algorithm 1+Algorithm 3 has a highest network throughput.

Fig. 8 plots the effect of the network throughput on different sub-channels, the number of devices is 30 and shows that it gradually increases with increasing in number of the sub-channels for Algorithm 4. This can be explained that the number of sub-channels is smaller than the number of SRs, we run the Algorithm 2 and Algorithm 3. With the increase number of the sub-channels, the number of sub-channels is larger than the number of SRs, we run the Algorithm 1 and Algorithm 3. we can make good use of communication resources under different sub-channels.

## V. CONCLUSION

In this paper, we present a cooperative federated learning framework for a MEC system with transmitting local models in a relay race manner, whose goal is maximizing the admission data to the edge server or the near devices. In CFLMEC, we use a decomposition approach to solve the problem by considering maximum data rate on sub-channel, channel reuse and wireless resource allocation in which establish a primal-dual learning framework and batch gradient decent to learn the dynamic network with outdated information and predict the sub-channel condition. With aim at maximizing throughput of devices, we propose communication resource allocation algorithms with and without sufficient sub-channels for strong reliance on edge servers (SRs) in cellular link, and interference aware communication resource allocation algorithm less reliance on edge servers (LRs) in D2D link. At the same time, we analyze the complexity of the proposed algorithms. Finally, we conduct extensive experiments to evaluate the performance of the CFLMEC and the results show that the proposed method can achieve a higher throughput compared with exiting work.